\documentclass[12pt]{article}
\usepackage{graphicx}
\usepackage{amsmath}
\usepackage{slashed}
\usepackage{mathrsfs}
\usepackage{epsfig}
\usepackage{lipsum}% just to generate text for the example
\usepackage{mathrsfs}
\usepackage{epsfig}
\usepackage{dcolumn}% Align table columns on decimal point
\usepackage{bm}% bold math
\usepackage{listings}
\def\beq{\begin{equation}}
\def\eeq{\end{equation}}
\def\bea{\begin{eqnarray}}
\def\eea{\end{eqnarray}}
\def\nn{\nonumber}
\def\nl{\nonumber \\}

\def\bra{\langle}
\def\ket{\rangle}

\usepackage{inputenc}

\def\bra#1{\left\langle #1\right|}

\def\lsim{\raise0.3ex\hbox{$\;<$\kern-0.75em\raise-1.1ex
\hbox{$\sim\;$}}}
\def\gsim{\raise0.3ex\hbox{$\;>$\kern-0.75em\raise-1.1ex
\hbox{$\sim\;$}}}

\newcommand{\be} {\begin{equation}}
\newcommand{\ee} {\end{equation}}
\newcommand{\ba} {\begin{eqnarray}}
\newcommand{\ea} {\end{eqnarray}}

\usepackage{graphicx}
\usepackage{amsthm}
\usepackage{amsmath}
\usepackage{graphicx,subfigure}% Include figure files
\usepackage{dcolumn}% Align table columns on decimal point
\usepackage{bm}% bold math
\usepackage{slashed}
\usepackage{calligra}
\usepackage{amsfonts}

\usepackage{amsmath,amssymb,bm,multirow,float,bbold,hyperref}
\usepackage{graphicx}
\usepackage[american]{babel}

\usepackage{enumerate}

\def \lb{\Lambda_b}
\def \lc{\Lambda_c}

\def\beq{\begin{equation}}
\def\eeq{\end{equation}}
\def\bea{\begin{eqnarray}}
\def\eea{\end{eqnarray}}
\def\ber{\begin{eqnarray*}}
\def\eer{\end{eqnarray*}}
\def\bwt{\begin{widetext}}
\def\ewt{\end{widetext}}
\def\nn{\nonumber}
\def\roughly#1{\mathrel{\raise.3ex\hbox
{$#1$\kern-.75em\lower1ex\hbox{$\sim$}}}}
\def\lsim{\roughly<}
\def\gsim{\roughly>}

\def\order{\lower 1.8ex \hbox{\LARGE\~{}}}

\def\bra#1{\left\langle #1\right|}
\def\ket#1{\left| #1\right\rangle}
\def \({\left(}
\def \){\right)}
\def \[{\left[}
\def \]{\right]}
\def \l|{\left|}
\def \r|{\right|}

\def \nn{\nonumber}
\def \nl{\nn \\}

\def \be{\beta}

\def \lb{\Lambda_b}
\def \lc{\Lambda_c}

\def\order{\lower 1.8ex \hbox{\LARGE\~{}}}

\begin{document}

\unitlength = 1mm
\begin{flushright}
SHIP-HEP-2020-02 \\
[10mm]
\end{flushright}

\begin{center}
\bigskip {\Large  \bf Reading the Lattice QCD Form Factors of the $\Lambda_b \rightarrow \Lambda_c$ Transition Using a C-Code}
\\[8mm]
Bradley Smith  $^{\dag}$ 
\footnote{E-mail:
\texttt{bs5797@cs.ship.edu}} 
and Ahmed Rashed $^{\dag}$  
\footnote{E-mail:
\texttt{amrashed@ship.edu}} 
\\[3mm]
\end{center}

\begin{center}
~~~{\it $^{\dag}$ Department  of Physics,}\\ 
~~~{ \it Shippensburg University of Pennsylvania,}\\
~~~{\it  Franklin Science Center,} \\
~~~{\it 1871 Old Main Drive, Pennsylvania, 17257, USA}\\
\end{center}

%\newpage

\begin{center} 
\bigskip (\today) \vskip0.5cm {\Large Abstract\\} \vskip3truemm
\parbox[t]{\textwidth}  
{
The $\Lambda_b \rightarrow \Lambda_c \ell^- \bar{\nu}_\ell$ process has been discussed in the literatures as a probe for new physics. This process receives contributions in new physics models from scalar, vector, and tensor hadronic currents. The form factors of these currents can be obtained from the quark model or the lattice QCD. In this work, we present a c-code to read the scalar, vector, and tensor lattice QCD form factors. A Mathematica code was used in Ref.~\cite{Datta:2017aue} to read the form factors. Our c-code is much faster than the Mathematica code where the ratio of the wall clock time for the c-code compared to the Mathematica code is 1:64.2 per data point.

 }
\end{center}

\thispagestyle{empty} \newpage \setcounter{page}{1}
% Decrease texheight (for preprint numbers) again
%\textheight 23.0 true cm
\baselineskip=14pt

\section{Program Summary}

\begin{itemize}
\item \textbf{Program title:} Lattice$\_$QCD

\item \textbf{Licensing provisions:} GNU General Public License 3 (GPL)

\item \textbf{Programming language:} c

\item \textbf{Supplementary material (if any):} 

\begin{enumerate}
\item \textit{Main Code:}  main.c
\item \textit{Supplementary Codes:} Constant.c, Constant.h, Calculate.c, Calculate.h, File$\_$Manipulate.c, File$\_$Manipulate.h
\item \textit{Data Files:} covariance, results, HO$\_$covariance, HO$\_$results
\end{enumerate}

%\item Journal Reference of previous version:**

%\item Does the new version supersede the previous version?:**

%\item Reasons for the new version:**

%\item Summary of revisions:**

\item \textbf{Nature of problem:} In this paper, we present a code in the c programming language to read the scalar, vector, and tensor lattice QCD form factors of the transition $\Lambda_b \to \Lambda_c$ and generate data points for the differential decay rate of the process $\Lambda_b \to \Lambda_c \tau \bar{\nu}_\tau$. We convert the Mathematica code written by (SM) and used in Ref.~\cite{Datta:2017aue} into a c-code. The code is verified by comparing its results with the results of the Mathematica code where we found that the ratio of the wall clock time for the c-code compared to the Mathematica code is 1:64.2 per data point. The advantage of the c-code over the Mathematica code is in its efficiency, both in terms of speed and memory. The c language is a lower level language and thus has less overhead, this enables the creation of smaller, faster programs with higher flexibility and more control over available computer resources.

\item \textbf{Solution method:}   The code was broken down into three core sections, each with its own challenges in its conversion. The first section is reading in the files. For this section we had to write four read file codes to read in the data from four different files. After the data was read in it was in an undesirable order, so it had to be reorganized into the desired order for the code to run smoothly. 
  The next section was the calculations which proved difficult as we had to map out the order of data for the code. The data flow of the  Mathematica code was not initially clear. This was solved by working backwards from the solution, following the references used in the Mathematica functions to find the dependencies and flow of data. Many of the functions in Mathematica do not exist in c, this meant that many of the functions used in the Mathematica code had to be remade in the c-code. Many patterns where identified to simplify the code, such as the calculations for finding the nominal and higher order form factors where similar in form though they did need alterations to their functions. The formula of the decay rates is very large and complex and thus needed to be broken down. 
  The final section was the data points for making the graphs which are organized into a table that compares the desired information; this table is then placed into a text file for easy transportation into a graphing program like Excel. The table contains the $q^2$, the differential decay rate and the pos/neg errors. The differential decay rate is plotted versus $q^2$.

\item \textbf{Additional comments including Restrictions and Unusual features:}

\item \textbf{References:} Refer to section \ref{RefList}

\item \textbf{Code url:} \url{https://github.com/darkfiresmith96/Lattice_QCD}

\end{itemize}

\section{Introduction}

Finding new physics beyond the Standard Model (SM) has been the cornerstone of modern particle physics. Studying the processes where colliders may have an access provides an opportunity to search for new physics (NP). For instance, $\Lambda_b \to \Lambda_c \ell^- \bar{\nu}_\ell$ can be used to test the limits of the Non-Relativistic Quark (NRQ) model calculations \cite{Chakraverty:1996sb,Albertus:2004wj}. It also can confirm possible new physics and  point to the correct model of new physics that explain the $R(D^{(*)})$ and $R(K^{(*)})$ puzzles \cite{Datta:2017aue, Shivashankara:2015cta}, where in the processes of $\bar{B} \to D^{(*)}\ell \nu_\ell$ \cite{Lees:2013uzd,Less:2013uzd2,Bhattacharya:2014wla, Bhattacharya:2016mcc} and $B \to K^{(*)} \ell^+ \ell^-$ \cite{Aaij:2014ora,Bhattacharya:2014wla, Bhattacharya:2016mcc} the experimental results have shown that there is universality violation of $(\tau, \mu)$ and $(\mu, e)$, consequently. This could be a clear sign of  new physics   involving either new heavy \cite{Alok:2017sui, Datta:2019zca}  or light states \cite{Datta:2017pfz,Datta:2017ezo}.
On the other hand, measuring the decay rates of the $\Lambda_b \to p \,\ell^- \bar{\nu}_\ell$ and $\Lambda_b \to \Lambda_c \,\ell^- \bar{\nu}_\ell$ is useful in the determination of the CKM matrix element $V_{ub}$ and $V_{cb}$ \cite{Datta:1995mv}. Therefore, precise determination of the form factor is necessary in calculating accurate $V_{ub}$ and $V_{cb}$. For that, the hadronic transition $\Lambda_b \to \Lambda_c$ is useful in searching for new physics beyond the Standard Model.

%Lepton universality in the Standard Model is a main property in flavor physics. Recently, experimental evidences for the violation of the lepton universality have been observed in different decay channels. 

%The decay $\Lambda_b \to \Lambda_c \ell^-\bar{\nu}_\ell$has not been measured experimentally though it might be possible to observe this decay at the LHCb. 

Recently, the LHCb Collaboration \cite{Aaij:2015bfa} reported the first measurement 
of the ratio of the heavy-to-light semileptonic $\Lambda_b \to p \,\ell^- \bar{\nu}_\ell$  and heavy-to-heavy semileptonic $\Lambda_b \to \Lambda_c \,\ell^- \bar{\nu}_\ell$  decay rates in the constrained kinematical regions.
The hadronic current of the transition $\Lambda_b \to \Lambda_c$ can be parameterized in terms of scalar, vector, and tensor form factors. There are two ways of obtaining the form factors; the quark model \cite{Cardarelli:1997sx,Dosch:1997zx,Huang:1998rq,MarquesdeCarvalho:1999bqs,Huang:2004vf,Pervin:2005ve,Ke:2007tg,Wang:2009hra,Azizi:2009wn,Khodjamirian:2011jp,Gutsche:2014zna} and the lattice QCD \cite{Detmold:2015aaa, Datta:2017aue}. 
 Using these form factors, one can present predictions for the
$\Lambda_b \to p \,\ell^- \bar{\nu}_\ell$ and $\Lambda_b \to \Lambda_c \,\ell^- \bar{\nu}_\ell$ differential and integrated decay rates.

In Ref. \cite{Datta:2017aue}, the lattice QCD form factors of the $\Lambda_b \to \Lambda_c$ have been used for the scalar, vector, and tensor hadronic currents. In Ref.~\cite{Detmold:2015aaa}, the authors have used the lattice QCD with 2+1 flavors of dynamical domain-wall fermions in calculating the form factors. One of the authors (Stefan Meinel) has extended the analysis of Ref.~\cite{Detmold:2015aaa} to include the tensor form factors defined in Ref. \cite{Datta:2017aue}. The tensor form factors were extracted from the lattice QCD correlation functions using ratios defined as in Ref.~\cite{Detmold:2016pkz}. The central values and statistical uncertainties of the form factors (and of any observables depending
on the form factors) were evaluated using the "nominal fit". On the other hand, the combined systematic uncertainty associated with the continuum extrapolation,
chiral extrapolation, z expansion, renormalization, scale setting, b-quark parameter tuning, finite volume, and missing isospin symmetry breaking/QED were evaluated using  the "higher-order fit" in conjunction with the nominal fit. The authors used a Mathematica code written by (SM) in reading the form factors. The full covariance matrices of the nominal and higher-order parameters of all
ten $\Lambda_b \to \Lambda_c$ form factors (vector, axial vector, and tensor) are provided in supplemental data files.

%  The $q^2$ will be the x-axis for the graph and the differential decay rate and errors will be on the y-axis.

The paper is organized in the following manner: In sec. (2) we introduce the effective Lagrangian to parametrize the NP operators of the transition  $b\to c\tau^-\bar{\nu}_\tau$  where the hadronic currents are defined in terms of the QCD lattice scalar, vector, and tensor form factors. In sec. (3) we present the c-code that reads the lattice form factors of the transition $\Lambda_b \rightarrow \Lambda_c$ and generate data. Discussion for the results and the future of the code are given in sec. (4, 5), consequently. We conclude in sec. (6).

\section{Lattice Form Factor}

In the presence of NP, the effective Hamiltonian for the quark-level transition $b\to c\tau^-\bar{\nu}_\tau$  can be written in the form \cite{Chen:2005gr,Bhattacharya:2011qm,Datta:2012qk}
\bea
\label{eq1:Lag}
 {\cal{H}}_{eff} &=&  \frac{G_F V_{cb}}{\sqrt{2}}\Big\{
\Big[\bar{c} \gamma_\mu (1-\gamma_5) b  + g_L \bar{c} \gamma_\mu (1-\gamma_5)  b + g_R \bar{c} \gamma_\mu (1+\gamma_5) b\Big] \bar{\tau} \gamma^\mu(1-\gamma_5) \nu_{\tau} \nl && +  \Big[g_S\bar{c}  b   + g_P \bar{c} \gamma_5 b\Big] \bar{\tau} (1-\gamma_5)\nu_{\tau} + \Big[g_T\bar{c}\sigma^{\mu \nu}(1-\gamma_5)b\Big]\bar{\tau}\sigma_{\mu \nu}(1-\gamma_5)\nu_{\tau} + h.c \Big\}, \nonumber \\ \label{eq:Heff}
\eea 
where  $G_F$ is the Fermi constant, $V_{cb}$ is the Cabibbo-Kobayashi-Maskawa (CKM) matrix element, and we use $\sigma_{\mu \nu} = i[\gamma_\mu, \gamma_\nu]/2$. We consider that the above Hamiltonian is written at the $m_b$ energy scale. 

%%Therefore, there is no need for running the couplings down from the high NP energy scale. Running the couplings in the case of vector current has been discussed in Refs.~(\cite{Feruglio:2016gvd}, \cite{Feruglio:2017rjo}).''
%If the effective interaction is written at the cut-ff scale $\Lambda$ then running down to the $m_b$ scale will generate new operators and new contributions,
%which have been discussed in Refs.~\cite{Feruglio:2016gvd, Feruglio:2017rjo}. These new contributions can strongly constrain models but to really calculate their true impacts we have to consider specific models
% where there might be cancellations between various terms.
%%Therefore, there is no need for running the couplings down from the high NP energy scale.  Running the couplings in the case of vector current has been discussed in Refs.~(\cite{Feruglio:2016gvd}, \cite{Feruglio:2017rjo}).

%The SM effective Hamiltonian corresponds to $g_L = g_R = g_S = g_P = g_T = 0$.
%In Eq.~(\ref{eq:Heff}), we have assumed the neutrinos to be always left chiral. In general, with NP the neutrino associated with the $\tau$ lepton does not have to carry the same flavor. 

The hadronic helicity amplitudes of the transition $\Lambda_{b}(p_{\lb})\rightarrow\Lambda_{c}(p_{\lc})$ can be expressed, according to the Lagrangian above, in terms of the scalar-type, vector/axial-vector-type, and tensor-type form factors as
\bea
H^{SP}_{\lambda_{\Lambda_c},\lambda=0}&=&H^S_{\lambda_{\Lambda_c},\lambda=0}+H^P_{\lambda_{\Lambda_c},\lambda=0}, \nn\\
H^S_{\lambda_{\Lambda_c},\lambda=0}&=&g_S \bra{\lc}\bar{c} b\ket{\lb},\nn\\
H^P_{\lambda_{\Lambda_c},\lambda=0}&=&g_P \bra{\lc}\bar{c}\gamma_5 b\ket{\lb},
\eea
\bea
H^{VA}_{\lambda_{\Lambda_c},\lambda}&=&H^V_{\lambda_{\Lambda_c},\lambda}-H^A_{\lambda_{\Lambda_c},\lambda}, \nn\\
H^V_{\lambda_{\Lambda_c},\lambda}&=&(1+g_L+g_R)\,\epsilon^{*\mu}(\lambda)\bra{\lc}\bar{c}\gamma_{\mu} b\ket{\lb}, \nn\\
H^A_{\lambda_{\Lambda_c},\lambda}&=&(1+g_L-g_R)\,\epsilon^{*\mu}(\lambda)\bra{\lc}\bar{c}\gamma_{\mu}\gamma_5 b\ket{\lb},
\eea
and
\bea
H^{(T){\lambda_{\lb}}}_{\lambda_{\lc},\lambda ,\lambda^{\prime}}&=&H^{(T1){\lambda_{\lb}}}_{\lambda_{\lc},\lambda ,\lambda^{\prime}}-H^{(T2){\lambda_{\lb}}}_{\lambda_{\lc},\lambda ,\lambda^{\prime}}, \nn\\
H^{(T1){\lambda_{\lb}}}_{\lambda_{\lc},\lambda ,\lambda^{\prime}}&=&g_T\: \epsilon^{*\mu}(\lambda)\epsilon^{*\nu}(\lambda^{\prime})\bra{\lc}\bar{c}i\sigma_{\mu \nu} b\ket{\lb},\nn\\
H^{(T2){\lambda_{\lb}}}_{\lambda_{\lc},\lambda ,\lambda^{\prime}}&=&g_T\:\epsilon^{*\mu}(\lambda)\epsilon^{*\nu}(\lambda^{\prime})\bra{\lc}\bar{c}i\sigma_{\mu \nu}\gamma_5 b\ket{\lb}.
\eea
%
%The leptonic amplitudes are defined as
%\bea
%L^{\lambda_\tau}&=&\bra{\tau\bar{\nu}_\tau}\bar{\tau} (1-\gamma_5)\nu_\tau\ket{0}, \nn\\
%L^{\lambda_\tau}_{\lambda}&=&\epsilon^\mu (\lambda)\bra{\tau\bar{\nu}_\tau}\bar{\tau}\gamma_\mu (1-\gamma_5)\nu_\tau\ket{0}, \nn\\
%L^{\lambda_\tau}_{\lambda ,\lambda^{\prime}}&=&-i\epsilon^\mu (\lambda)\epsilon^\nu (\lambda^\prime)\bra{\tau\bar{\nu}_\tau}\bar{\tau}\sigma_{\mu \nu} (1-\gamma_5)\nu_\tau\ket{0}.
%\eea
%Above, $\epsilon^{\mu}$ are the polarization vectors of the virtual vector boson (see Appendix \ref{sec:spinorsandvectors}). The explicit expressions for the hadronic and leptonic helicity amplitudes are presented in the following.

%\subsubsection{Hadronic helicity amplitudes}

In the amplitudes above, we use the helicity-based definition of the $\Lambda_b\to\Lambda_c$ form factors, which was introduced in \cite{Feldmann:2011xf}. 
The matrix elements of the vector and axial vector currents can be written in terms of six helicity form factors $F_+$, $F_\perp$, $F_0$, $G_+$, $G_\perp$, and $G_0$
as follows:
\bea
\bra{\lc}\bar{c}\gamma^\mu b\ket{\lb}&=&\bar{u}_{\lc}\Big[ F_0 (q^2)(m_{\lb} - m_{\lc})\frac{q^\mu}{q^2} \nonumber\\
&&+F_+ (q^2)\frac{m_{\lb} + m_{\lc}}{Q_+}(p_{\lb}^{\mu} +p_{\lc}^{\mu}-(m_{\lb} ^2 - m_{\lc} ^2)\frac{q^\mu}{q^2}) \nonumber\\
&&+F_\perp (q^2)(\gamma^\mu - \frac{2m_{\lc}}{Q_+}p_{\lb}^{\mu} - \frac{2m_{\lb}}{Q_+}p_{\lc}^{\mu})\Big]u_{\lb}, \label{eq:VFF} \\
\bra{\lc}\bar{c}\gamma^\mu \gamma_5 b\ket{\lb}&=&-\bar{u}_{\lc}\gamma_5\Big[ G_0 (q^2)(m_{\lb} + m_{\lc})\frac{q^\mu}{q^2} \nonumber\\
&&+G_+ (q^2)\frac{m_{\lb} - m_{\lc}}{Q_-}(p_{\lb}^{\mu} +p_{\lc}^{\mu}-(m_{\lb} ^2 - m_{\lc} ^2)\frac{q^\mu}{q^2})\nonumber\\
&&+G_\perp (q^2)(\gamma^\mu + \frac{2m_{\lc}}{Q_-}p_{\lb}^{\mu} - \frac{2m_{\lb}}{Q_-}p_{\lc}^{\mu})\Big]u_{\lb} \label{eq:AFF}.
\eea
The matrix elements of the scalar and pseudoscalar currents can be obtained from the vector and axial vector matrix elements using the equations of motion:
\bea
\nonumber \bra{\lc}\bar{c} b\ket{\lb} &=& \frac{q_\mu}{m_b-m_c}\bra{\lc}\bar{c}\gamma^\mu b\ket{\lb} \\
&=& F_0(q^2)  \frac{m_{\lb} - m_{\lc}}{m_b-m_c} \bar{u}_{\lc}u_{\lb}, \\
\nonumber \bra{\lc}\bar{c}\gamma_5 b\ket{\lb} &=& \frac{q_\mu}{m_b+m_c}\bra{\lc}\bar{c}\gamma^\mu\gamma_5 b\ket{\lb} \\
&=& G_0(q^2)  \frac{m_{\lb} + m_{\lc}}{m_b+m_c} \bar{u}_{\lc}\gamma_5 u_{\lb}.
\eea
The matrix elements of the tensor currents can be written in terms of four form factors $h_+$, $h_\perp$, $\widetilde{h}_+$, $\widetilde{h}_\perp$,
\bea
&&\bra{\lc}\bar{c}i\sigma^{\mu\nu} b\ket{\lb}=\bar{u}_{\lc}\Big[2h_+(q^2)\frac{p_{\lb}^\mu p_{\lc}^{ \nu}-p_{\lb}^\nu p_{\lc}^{\mu}}{Q_+} \nonumber\\
&&+h_\perp (q^2)\Big(\frac{m_{\lb}+m_{\lc}}{q^2}(q^\mu \gamma^\nu -q^\nu \gamma^\mu)-2(\frac{1}{q^2}+\frac{1}{Q_+})(p_{\lb}^\mu p_{\lc}^{\nu}-p_{\lb}^\nu p_{\lc}^{\mu}) \Big) \nonumber\\
&&+\widetilde{h}_+ (q^2)\Big(i\sigma^{\mu \nu}-\frac{2}{Q_-}(m_{\lb}(p_{\lc}^{\mu}\gamma^\nu -p_{\lc}^{\nu}\gamma^\mu)\nonumber\\
&&-m_{\lc}(p_{\lb}^\mu \gamma^\nu -p_{\lb}^\nu \gamma^\mu)+p_{\lb}^\mu p_{\lc}^{\nu}-p_{\lb}^\nu p_{\lc}^{\mu}) \Big) \nonumber\\
&&+\widetilde{h}_\perp(q^2) \frac{m_{\lb}-m_{\lc}}{q^2 Q_-}\Big((m_{\lb}^2-m_{\lc}^2-q^2)(\gamma^\mu p_{\lb}^\nu - \gamma^\nu p_{\lb}^\mu)\nonumber\\
&&-(m_{\lb}^2-m_{\lc}^2+q^2)(\gamma^\mu p_{\lc}^{\nu}-\gamma^\nu p_{\lc}^{\mu})+2(m_{\lb}-m_{\lc})(p_{\lb}^\mu p_{\lc}^{\nu}-p_{\lb}^\nu p_{\lc}^{\mu}) \Big)
\Big]u_{\lb}. \nonumber \\ \label{eq:TFF}
\eea
%The matrix elements of the current $\bar{c}i\sigma^{\mu\nu}\gamma_5b$ can be obtained from the above equation by using the identity
%\bea
%\sigma^{\mu \nu}\gamma_{5}=-\frac{i}{2}\epsilon^{\mu \nu \alpha \beta}\sigma_{\alpha \beta}.
%\eea 
%

The full covariance matrices of the nominal and higher-order parameters of all ten $\Lambda_b \rightarrow \Lambda_c$ form factors (vector, axial vector, and tensor) are provided as supplemental files. In the following section, we will provide the c-code which will be used in reading the files and generating data points. The differential decay rate of the process $\Lambda_b \to \Lambda_c \,\ell^- \bar{\nu}_\ell$ is given in Ref.~\cite{Datta:2017aue}. The c-code will be used to provide the differential decay rate graph for the SM and NP at $g_s= g_p= g_R= g_T=0$ and $g_L=-2.2$. This will be compared with the result from the Mathematica code.

\section{Code}
The code is split up into 3 parts: the constants part, the data manipulation part, and the calculations part.  These 3 parts each have their own file and are brought together by the main file. At the start of the code is the constants part, which declares and stores each of the constants usable in the larger code.  The constants file is the following:
\begin{lstlisting}
//constants used
/**
 *  the number of data points generated
 */
const int points = 10;
/**
 * the starting point for the data
 */
double start = 0;
/**
 *  the ending point for the data
 */
double end = 0;
/**
 *some important constants
 */
const double mlambdab = 5.61951;
const double mlambdac = 2.28646;
const double GF = 1.1663787 * pow(10, -5);
const double hbar = 6.582119514 * pow(10,-25);
const double mtau = 1.77682;
const double mmu = 0.105658;
const double PI = 3.14159265359;
const double vbc = 0.0414;
const double vcb = 0.0414;
const double m1 = 5.6195;
const double m2 = 2.28646;
const double mb = 4.18;
const double mc = 1.257;
const double eps = 1/pow(10, 8);
const double epsneg = -1/pow(10,8);
/**
 * pole locations
 */
const double mb0minus = 6.276;
const double mb1minus = 6.332;
const double mb0plus = 6.725;
const double mb1plus = 6.768;
/**
 * the number of parameters for the nominal data set
 */
const int set1 = 18;
/**
 *  the number of parameters for the higher order data set
 */
const int set2 = 28;
/**
 *  the number of form factors
 */
const int numfactors = 10;
    //t0 parameter of z expansion
double tminus = 0;
double t0 = 0;
double qsqrmax = 0;

/**
 *  the names of the parameters for the nominal form factors
 */
const char *nominalparam[] = {"a0_fplus", "a1_fplus", "a0_fperp",
 "a1_fperp", "a0_f0", "a1_f0",
 "a0_gpp", "a1_gplus", "a1_gperp", "a0_g0", "a1_g0", "a0_hplus",
 "a1_hplus", "a0_hperp", "a1_hperp", "a0_htildepp", "a1_htildeplus",
 "a1_htildeperp"};
/**
 *  the names of the parameters for the higher order form factors
 */
const char *hoparam[] = {"a0_fplus", "a1_fplus", "a2_fplus", 
  "a0_fperp", "a1_fperp",
  "a2_fperp", "a0_f0", "a1_f0", "a2_f0", "a0_gpp", "a1_gplus",
  "a2_gplus", "a1_gperp", "a2_gperp", "a0_g0", "a1_g0", "a2_g0",
  "a0_hplus", "a1_hplus", "a2_hplus", "a0_hperp", "a1_hperp",
  "a2_hperp", "a0_htildepp", "a1_htildeplus", "a2_htildeplus",
  "a1_htildeperp", "a2_htildeperp"};
/**
 *  the maximum length for file names
 */
int file_name_length = 300;
/**
 *  the number of parameters for the current data set
 */
int param_num = 0;
/**
 *  the precision, number of digits, to take in for the parameters
 */
int pressision = 30;
/**
 * the factors used
 */
double gs = 0;
double gp = 0;
double gL = 0;
double gR = 0;
double gT = 0;
double ml = mtau;
\end{lstlisting}
The main function starts with some final adjustments to the constants, due to the need for quick calculations in their stored value. The constants that are adjusted, as well as their new values are the following:
\begin{lstlisting}
	 /// finishes the declaration of some important variables
    tminus = pow(mlambdab - mlambdac, 2);
    t0 = tminus;
    qsqrmax = tminus;
    end = qsqrmax;
    start = qsqrmin(ml);
\end{lstlisting}
The code then takes in the file names of the files that contain the form factor parameters, their covariances, and their higher order forms. It then checks if they exist. If the files exist, then the code opens the four files. When the file names are typed in they need to be either in the same folder as the executable or include the path to the file.  The file name also requires its suffix to be opened.  once the data in the files are taken in and organized the files are closed. The code for taking in the files as well as the parameters for the code are in the following segment:
\begin{lstlisting}
    //this sets up the files used in the code
   /// For the nominal form factor parameters
    FILE *results_ptr1;
    char results_name1[file_name_length];
    /// For the nominal form factor covariances
    FILE *covariance_ptr1;
    char covariance_name1[file_name_length];
    /// For the higher order form factor parameters
    FILE *results_ptr2;
    char results_name2[file_name_length];
    /// For the higher order form factor covariances
    FILE *covariance_ptr2;
    char covariance_name2[file_name_length];
    /**
     *  Acquires the filename of the file with the nominal form 
     *  factor parameters from the user
     */
    printf("Enter in the filename of the file with the form factor 
      parameters for the first set: ");
    scanf("%s", results_name1);
    /**
     *  Opens the file with the nominal form factor parameters, ends 
     *  program if file isn't readable
     */
    results_ptr1 = fopen(results_name1, "r");
    if(results_ptr1 == NULL){
        printf("file cannot be opened, please check if it is in 
          correct location");
        exit(0);
    }
    /**
     *  Acquires the filename of the file with the nominal form 
     *  factor covariances from the user
     */
    printf("Enter in the file name of the file with the covariances 
      for the first set: "); //asks for file name
    scanf("%s", covariance_name1); //reads in file name
    /**
     *  Opens the file with the nominal form factor covariances, ends 
     *  program if file isn't readable
     */
    covariance_ptr1 = fopen(covariance_name1, "r");
    if(covariance_ptr1 == NULL){
        printf("file cannot be opened, please check if it is in 
          correct location");
        exit(0);
    }
    /**
     *  Acquires the filename of the file with the higher order form 
     *  factor parameters from the user
     */
    printf("Enter in the filename of the file with the form factor 
      parameters for the second set: "); //asks for file name
    scanf("%s", results_name2); //reads in file name
    /// Prints out file name, comment out
    //printf("%s\n", results_name);
    /**
     *  Opens the file with the higher order form factor parameters, 
     *  ends program if file isn't readable
     */
    results_ptr2 = fopen(results_name2, "r");
    if(results_ptr2 == NULL){
        printf("file cannot be opened, please check if it is in 
          correct location");
        exit(0);
    }
    /**
     *  Acquires the filename of the file with the higher order form 
     *  factor covariances from the user
     */
    printf("Enter in the file name of the file with the covariances 
      for the second set: "); //asks for file name
    scanf("%s", covariance_name2); //reads in file name
    /// Prints out file name, comment out
    //printf("%s\n", covariance_name);
    /**
     *  Opens the file with the higher order form factor covariances, 
     *  ends program if file isn't readable
     */
    covariance_ptr2 = fopen(covariance_name2, "r");
    if(covariance_ptr2 == NULL){
        printf("file cannot be opened, please check if it is in 
          correct location");
        exit(0);
    }
    // this takes in the additional variables
    printf("please enter in the gs: ");
    scanf("%lf", &gs);
    printf("please enter in the gp: ");
    scanf("%lf", &gp);
    printf("please enter in the gL: ");
    scanf("%lf", &gL);
    printf("please enter in the gR: ");
    scanf("%lf", &gR);
    printf ("please enter in the gT: ");
    scanf("%lf", &gT);
    /**
     *  Organize the parameters and covariances for the nominal form 
     *  factors into a usable order
    */
    param_num = set1;
    double parameters1[param_num];
    double covariances1[param_num][param_num];
    param_org(results_ptr1, parameters1, nominalparam);
    covar_org(covariance_ptr1, (double *)covariances1, nominalparam);
    /**
     *  Organizes the parameters and covariances for the nominal form 
     *  factors into a usable order
    */
    param_num = set2;
    double parameters2[param_num];
    double covariances2[param_num][param_num];
    param_org(results_ptr2, parameters2, hoparam);
    covar_org(covariance_ptr2, (double *)covariances2, hoparam);
    /**
     *  Closes the files, they are no longer needed
     */
    fclose(results_ptr1);
    fclose(results_ptr2);
    fclose(covariance_ptr1);
    fclose(covariance_ptr2);
}
\end{lstlisting}
The segment of code above contains 2 data manipulation functions, the first of which is param$\_$org.  This function takes in form factor variables from the files containing the r variables and organizes them in a form usable for the calculations. To organize the variables, it takes in the name of the variable and matches it up to the order stored in nomparam for the nominal parameters and hoparam for the higher order parameters. This function is showed in the following code segment.
\begin{lstlisting}
void param_org(FILE *file_ptr, double *param, const char 
  *factorlist[]){
    /// Used to check which array to place data; 0 read in parameter 
    /// name, 1 read in parameter
    int check = 0;
    /// Used to read in the chars
    char ch;
    /**
     *  Takes in the data and places it in the correct part of the 
     *  array
     */
    for(int i = 0; i < param_num; i++){
        /// Starts extraction of next letter
        ch = fgetc(file_ptr);
        /// the string version of the parameter
        char num[pressision];
        /// Stores the name of the variable being extracted
        char factor[file_name_length];
        /// Stores the position of the array that the respective char 
        /// will be stored on
        int k = 0;
        /// Extracts the variable and variable name
        while (ch != '\n'){
            /// Reads in the parameter name
            if (check == 0){
                /// Time to read in the parameter
                if(ch == ' '){
                    check = 1;
                    factor[k] = '\0';
                    k = 0;
                /// Read in next char of parameter name
                } else {
                    factor[k] = ch;
                    k++;

                }
            /// Reads in the parameter
            } else {
                num[k] = ch;
                k++;
            }
            /// Gets next char
            ch = fgetc(file_ptr);
        }
        /**
         *  Places the parameters into their correct positions in the 
         *  array
         */
        for(int j = 0; j < param_num; j++){
            /// Finds the array position using the parameter name
            if(strcmp(factor, factorlist[j]) == 0){
                /// Places parameter in found position as a double
                *(param + j) = atof(num); position
            }else{
                
            }
        }
        /// get next parameter
        check = 0;
    }
}
\end{lstlisting}
The second data manipulation function is covar$\_$org. This function takes in the form factor covariances from the files that contain the covariances and organizes them in a form usable for the calculations. To organize the covariances, it takes in the names of the two variables and maches them up to the rder stored in nomparam for the nominal parameters and hoparam for the higher order parameter. This function is showed in the following code segment:
\begin{lstlisting}
void covar_org(FILE *file_ptr, double *covar, const char 
  *factorlist[]){
    /// Used to check which array to place data; 0 read in parameter 
    /// 1 name, 1 read in parameter 2 name, 3 read in covariance 
    /// value
    int check = 0;
    /// Used to read in chars
    char ch;
    /**
     *  Takes in data and places it into the correct part of the 
     *  array
     */
    for(int i = 0; i < param_num; i++){
        for(int j = 0; j < param_num; j++){
            /// Start the extraction process
            ch = fgetc(file_ptr);
            /// Stores the string version of the covariance
            char num[pressision];
            /// Stores parameter 1 name
            char factor1[file_name_length];
            /// Stores parameter 2 name
            char factor2[file_name_length];
            ///Stores position of array char will be stored in
            int k = 0;
            /// Extracts covariance and parameter names
            while(ch != '\n'){
                /// Extract parameter 1 name
                if(check == 0){
                    /// Time to read in parameter 2 name
                    if(ch == ' '){
                        check = 1;
                        factor1[k] = '\0';
                        k = 0;
                        fgetc(file_ptr);
                    /// Read in next char of parameter 1 name
                    }else{
                        factor1[k] = ch;
                        k++;
                    }
                /// Extract parameter 2 name
                }else if(check == 1){
                    /// Time to read in covariance
                    if(ch == ' '){
                        check = 2;
                        factor2[k] = '\0';
                        k = 0;
                    /// Read in next char of parameter 2 name
                    }else{
                        factor2[k] = ch;
                        k++;
                    }
                /// Extract covariance
                }else {
                    num[k] = ch;//grabs number
                    k++;
                }
                /// Extract next char
                ch = fgetc(file_ptr);
            }
            /// Get next covariance
            check = 0;
            /**
             *  Places covariances into the correct part of the array
             */
            for(int x = 0; x < param_num; x++){
                /// Uses parameter 1 name to find position 1
                if(strcmp(factor1, factorlist[x]) == 0){
                    /// Uses parameter 2 name to find position 2
                    for(int y = 0; y < param_num; y++){
                        if(strcmp(factor2,factorlist[y]) == 0){
                            /// Places covariance in found position 
                            /// as double
                            *(covar + (x* param_num+y)) = atof(num);
                        }
                    }
                }
            }
        }
    }
}
\end{lstlisting}
Now, that the data is organized into their respective arrays and the files are properly closed, the code is ready to process the data and produce the decay rates and their errors for the data points.  The data points are produced using start, end, and points variables. The data is then pushed through a series of functions and simple calculations to produce the decay rates and the errors for each data point. The following segment shows the code used in the calculations:
\begin{lstlisting}
    /**
     *  Sets up the q^2 points, which acts as the independent 
     *  variable
     */
    double qsquared[points];
    qsquared[0] = start;
    for(int i = 1; i < points; i++){
        qsquared[i] = qsquared[i - 1] + ((end - start)/points);
    }
    /**
     *  Stores the decay rate, the dependent variable
     */
    double decay[points];
    /**
     *  Stores the error value for the decay rate
     */
    double error[points];
    /**
     *  Stores the values for the lower, "negative", error band
     */
    double negerror[points];
    /**
     *  Stores the values for the upper, "positive", error band
     */
    double poserror[points];
    /**
     *  This loop finds the decay and error values for each q^2 value
     */
    for(int i = 0; i <points; i++){
        /// Stores the nominal decay rate for the given q^2
        double resultnom = nominalcalc(qsquared[i], parameters1);
        /// Stores the higher order decay rate for the given q^2
        double resultHO = HOcalc(qsquared[i], parameters2);
        param_num = set1;
        /// Stores the nominal error for the given q^2
        double resulterrnom = resulterrorcalc(qsquared[i], resultnom, 
          parameters1, (double *)covariances1);
        param_num = set2;
        /// Stores the higher order error for the given q^2
        double resulterrHO = resulterrorcalc(qsquared[i], resultHO, 
          parameters2, (double *)covariances2);
        decay[i] = resultnom;
        error[i] = totalerrorcalc(resultnom, resultHO, resulterrnom, 
          resulterrHO);
        negerror[i] = decay[i] - error[i];
        poserror[i] = decay[i] + error[i];;
    }
\end{lstlisting}
The function "totalerrorcalc" takes in the values for the decay rate and higher order decay rate, as well as their individual errors, to find the total error in the calculation. It is the following segment:
\begin{lstlisting}
double totalerrorcalc(double resnom, double resHO, double errnom, 
  double errHO){
    /// error of the system
    double systerr;
    /// part 1 of error
    double systerr1;
    /// part 2 of error
    double systerr2;
    ///finds systerr1
    if(errHO > errnom){
        systerr1 = sqrt(pow(errHO, 2) - pow(errnom, 2));
    } else {
        systerr1 = 0;
    }
    ///finds systerr2
    systerr2 = resHO - resnom;
    if(systerr2 < 0){
        systerr2 = - systerr2;
    }
    ///finds systerr
    if(systerr1 >= systerr2){
        systerr = systerr1;
    } else {
        systerr = systerr2;
    }
    return sqrt(pow(errnom, 2) + pow(systerr, 2));
}
\end{lstlisting}
The function "resulterrorcalc" finds the individual errors for the decay rate and the higher order decay rate. It is in the following segment:
\begin{lstlisting}
double resulterrorcalc(double qsqr, double obs, double *param, double 
  *covar){
    /// cumulates the error
    double error = 0;
    ///finds error due to respective covariance
    double sum;
    for(int i = 0; i < param_num; i++){
        for(int j = 0; j < param_num; j++){
            sum = diffrentiate(qsqr, param, i) * *(covar+(i*param_num + j)) * diffrentiate(qsqr, param, j);
            error += sum;
        }
    }
    return sqrt(error);
}
\end{lstlisting}
The function "differentiate" uses numerical differentiation to find the derivative for the decay rate at the given point to be used in calculating the error. It is in the following segment:
\begin{lstlisting}
double diffrentiate(double qsqr, double *param, int num){
    double derivative;
    /// The next value for the decay
    double pos;
    /// The last value for the decay
    double neg;
    /// For the nominal factors
    if (param_num == set1){
        *(param + num) += eps;
        pos = nominalcalc(qsqr, param);
        *(param + num) -= 2 * eps;
        neg = nominalcalc(qsqr, param);
        *(param + num) += eps;
    /// For higher order factors
    }else {
        *(param + num) += eps;
        pos = HOcalc(qsqr, param);
        *(param + num) -= 2* eps;
        neg = HOcalc(qsqr, param);
        *(param + num)+= eps;
    }
    /// numerical differentiation
    derivative = (pos - neg)/(2 * eps) ;
    return derivative;
}
\end{lstlisting}
The function "nominalcalc" uses the form factor's parameters to find the form factors themselves. This function only processes the nominal form factors not the higher order factors. It is in the following segment:
\begin{lstlisting}
double nominalcalc(double qsqr, double param1[set1]){
    double nominal[numfactors];
    /// Used to keep track of the parameters used
    int counter1 = 0;
    /// Keeps track of array position
    int i = 0;
    /// Finds the nominal fplus
    nominal[i] = nomfactor(qsqr, mb1minus, param1[counter1], 
      param1[counter1 + 1]);
    counter1 += 2;
    i += 1;
    /// Finds the nominal fperp
    nominal[i] = nomfactor(qsqr, mb1minus, param1[counter1], 
      param1[counter1 + 1]);
    counter1 += 2;
    i += 1;
    /// Finds the nominal f0
    nominal[i] = nomfactor(qsqr, mb0plus, param1[counter1], 
      param1[counter1 + 1]);
    counter1 += 2;
    i += 1;
    /// Finds the nominal gplus
    nominal[i] = nomfactor(qsqr, mb1plus, param1[counter1], 
      param1[counter1 + 1]);
    i += 1;
    /// Finds the nominal gperp
    nominal[i] = nomfactor(qsqr, mb1plus, param1[counter1], 
      param1[counter1 + 2]);
    counter1 += 3;
    i += 1;
    /// Finds the nominal g0
    nominal[i] = nomfactor(qsqr, mb0minus, param1[counter1], 
      param1[counter1 + 1]);
    counter1 += 2;
    i += 1;
    /// Finds the nominal hplus
    nominal[i] = nomfactor(qsqr, mb1minus, param1[counter1], 
      param1[counter1 + 1]);
    counter1 += 2;
    i += 1;
    /// Finds the nominal hperp
    nominal[i] = nomfactor(qsqr, mb1minus, param1[counter1], 
      param1[counter1 + 1]);
    counter1 += 2;
    i += 1;
    /// Finds the nominal htildeplus
    nominal[i] = nomfactor(qsqr, mb1plus, param1[counter1], 
      param1[counter1 + 1]);
    i += 1;
    /// Finds the nominal htildeperp
    nominal[i] = nomfactor(qsqr, mb1plus, param1[counter1], 
      param1[counter1 + 2]);
    /// finds the nominal decay
    return dgammacalc(qsqr, nominal);
}
\end{lstlisting}
The function "HOcalc" also finds the form factors using their parameters, but for the higher order factors not the nominal factors. It is shown in the following segment:
\begin{lstlisting}
double HOcalc(double qsqr, double param2[set2]){
    double ho[numfactors];
    /// Keeps track of parameters used
    int counter1 = 0;
    /// Keeps track of array position
    int i = 0;
    /// Finds the higher order fplus
    ho[i] = highfactor(qsqr, mb1minus, param2[counter1], 
      param2[counter1 + 1], param2[counter1 + 2]);
    counter1 += 3;
    i += 1;
    /// Finds the higher order fperp
    ho[i] = highfactor(qsqr, mb1minus, param2[counter1], 
      param2[counter1 + 1], param2[counter1 + 2]);
    counter1 += 3;
    i += 1;
    /// Finds the higher order f0
    ho[i] = highfactor(qsqr, mb0plus, param2[counter1], 
      param2[counter1 + 1], param2[counter1 + 2]);
    counter1 += 3;
    i += 1;
    /// finds the higher order gplus
    ho[i] = highfactor(qsqr, mb1plus, param2[counter1], 
      param2[counter1 + 1], param2[counter1 + 2]);
    i += 1;
    /// Finds the higher order gperp
    ho[i] = highfactor(qsqr, mb1plus, param2[counter1], 
      param2[counter1 + 3], param2[counter1 + 4]);
    counter1 += 5;
    i += 1;
    /// Finds the higher order g0
    ho[i] = highfactor(qsqr, mb0minus, param2[counter1], 
      param2[counter1 + 1], param2[counter1 + 2]);
    counter1 += 3;
    i += 1;
    /// Finds the higher order hplus
    ho[i] = highfactor(qsqr, mb1minus, param2[counter1], 
      param2[counter1 + 1], param2[counter1 + 2]);
    counter1 += 3;
    i += 1;
    /// Finds the higher order hperp
    ho[i] = highfactor(qsqr, mb1minus, param2[counter1], 
      param2[counter1 + 1], param2[counter1 + 2]);
    counter1 += 3;
    i += 1;
    /// Finds the higher order htildeplus
    ho[i] = highfactor(qsqr, mb1plus, param2[counter1], 
      param2[counter1 + 1], param2[counter1 + 2]);
    i += 1;
    /// Finds the higher order htildeperp
    ho[i] = highfactor(qsqr, mb1plus, param2[counter1], 
      param2[counter1 + 3], param2[counter1 + 4]);
    /// finds the higher order decay
    return dgammacalc(qsqr, ho);
}
\end{lstlisting}
The function "dgammacalc" uses the form factors to solve the Hamiltonian and find the decay rates. The function is in the following segment:
\begin{lstlisting}
double dgammacalc(double qsqr, double factors[numfactors]){
    double rootbase = sqrt(pow(mlambdab,2) + pow(pow(mlambdac,2) - 
      qsqr,2)/pow(mlambdab,2)- 2*(pow(mlambdac,2)+qsqr));
    double root1 = sqrt(pow(mlambdab,2)-2*mlambdab*mlambdac
      +pow(mlambdac,2)-qsqr);
    double root2 = sqrt(pow(mlambdab+mlambdac,2)-qsqr);
    double root3 = sqrt(pow(mlambdab,2)+2*mlambdab*mlambdac
      +pow(mlambdac,2)-qsqr);
    double power1 = pow((1+gL)*(factors[1]*root1-factors[4]*root2),
      2);
    double f1 = 2*(pow(ml,2)+2*qsqr)*power1;
    double power2 = pow((1+gL)*(factors[1]*root1+factors[4]*root2),
      2);
    double f2 = 2 * (pow(ml,2)+2*qsqr) * power2;
    double power34 = pow(1/sqrt(qsqr)*(1+gL)*(factors[0]*(mlambdab + 
      mlambdac)*root1 + factors[3] * (mlambdac - mlambdab) * root3),
      2);
    double f3 = pow(ml,2)*power34;
    double f4 = 2*qsqr * power34;
    double power5 = pow(1/sqrt(qsqr)*(1+gL)*(factors[2]*(mlambdab - 
      mlambdac)*root2 - factors[5]*(mlambdab+mlambdac)*root1),2);
    double f5 =3*pow(ml,2)*power5;
    double power6 = pow(1/sqrt(qsqr)*(1+gL)*(factors[2]*(mlambdab - 
      mlambdac)*root2 + factors[5]*(mlambdab+mlambdac)*root1),2);
    double f6 = 3 * pow(ml,2)*power6;
    double power78 = pow(1/sqrt(qsqr)*(1+gL)*(factors[0]*(mlambdab
      +mlambdac)* root1+ factors[3]*(mlambdab-mlambdac)*root2),2);
    double f7 = pow(ml,2)*power78;
    double f8 = 2 * qsqr * power78;
    double sum = f1 + f2 + f3 + f4 + f5 + f6 + f7 + f8;
    double base = 1/(768* pow(m1,2)*pow(PI,3)* pow(qsqr,2))*pow(GF,
      2)*pow(pow(ml,2)-qsqr,2)*rootbase*pow(vbc,2)*sum * pow(10,15);
    return base;
}
\end{lstlisting}
 The functions in the following segment conduct small calculations that are used throughout the code:
\begin{lstlisting}
/**
 * finds the starting value for qsqr
 */
double qsqrmin (double ml){
    return pow(ml,2);
}
/**
 *  performs the zz function
 */
double zz(double qsqr, double tplus){
    double root1 = sqrt(tplus- qsqr);
    double root2 = sqrt(tplus - t0);
    double top = root1 - root2;
    double bottom = root1 + root2;
    double result = top/bottom;
    return result;
}
/**
 *  performs the pp function
 */
double pp (double qsqr, double mpole){
    double polesqr = pow(mpole,2);
    double result = 1 - qsqr/polesqr;
    return result;
}
/**
 *  finds the nominal form factors
 */
double nomfactor (double qsqr, double param1, double param2, double   
  param3){
    return 1/pp(qsqr, param1) * (param2 + param3 * zz(qsqr, 
      pow(param1, 2)));
}
/**
 *  finds the higher order form factors
 */
double highfactor (double qsqr, double param1, double param2, double 
  param3, double param4){
      double parampow = pow(param1, 2);
      double zzresult = zz(qsqr, parampow);
      double ppresult = pp(qsqr, param1);
      double factor = 1 / ppresult * (param2 + param3 * zzresult + 
        param4 * pow(zzresult,2));
      return factor;
}
/**
 *  finds the current splus value
 */
double splus(double qsqr){
    return pow(mlambdab + mlambdac, 2) - qsqr;
}
/**
 *  finds the current sminus value
 */
double sminus(double qsqr){
    return pow(mlambdab - mlambdac, 2) - qsqr;
}
\end{lstlisting}
Once the decay and error for each qsqr value is found the code then places the data into a file. The file is organized as a data table for easy transition into a graphing software.  The code that sets up the data file is in the following segment:
\begin{lstlisting}
/// The filename of the file used to store the data
    char datafilename[] = "data.txt";
    /// Sets up the file for reading in the data, if file exists it 
    /// clears it out
    FILE *datafile = fopen(datafilename, "w");
    fprintf(datafile, "qsqr decay negerror poserror\n");
    /// Places the data into the file
    for(int i = 0; i < points; i++){
        fprintf(datafile, "%lf %lf %lf %lf\n", qsquared[i], 
          decay[i], negerror[i], poserror[i]);
    }
    /// closes the file, it is finished
    fclose(datafile);
\end{lstlisting}

\section{Results}

\begin{figure}[h]
 \centering
 \includegraphics{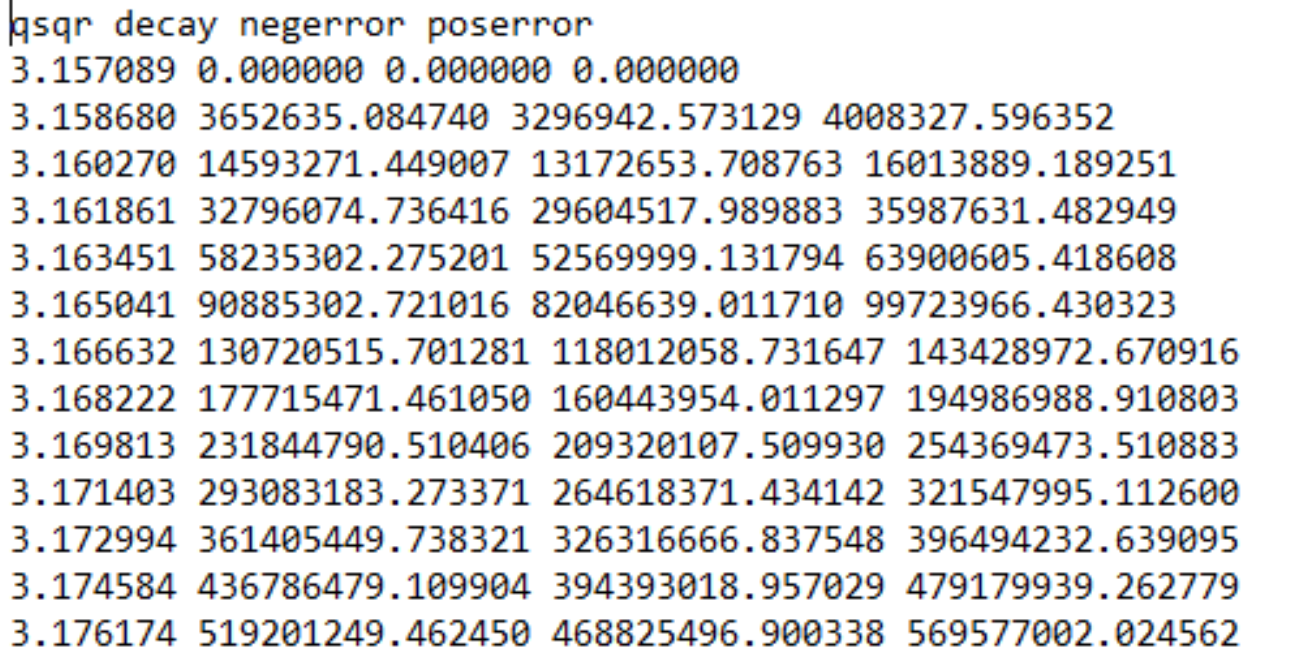}
 \caption{Example of data from code}
 \label{figure1}
\end{figure}

\begin{figure}[!h]
 \centering
 \includegraphics[scale=0.55]{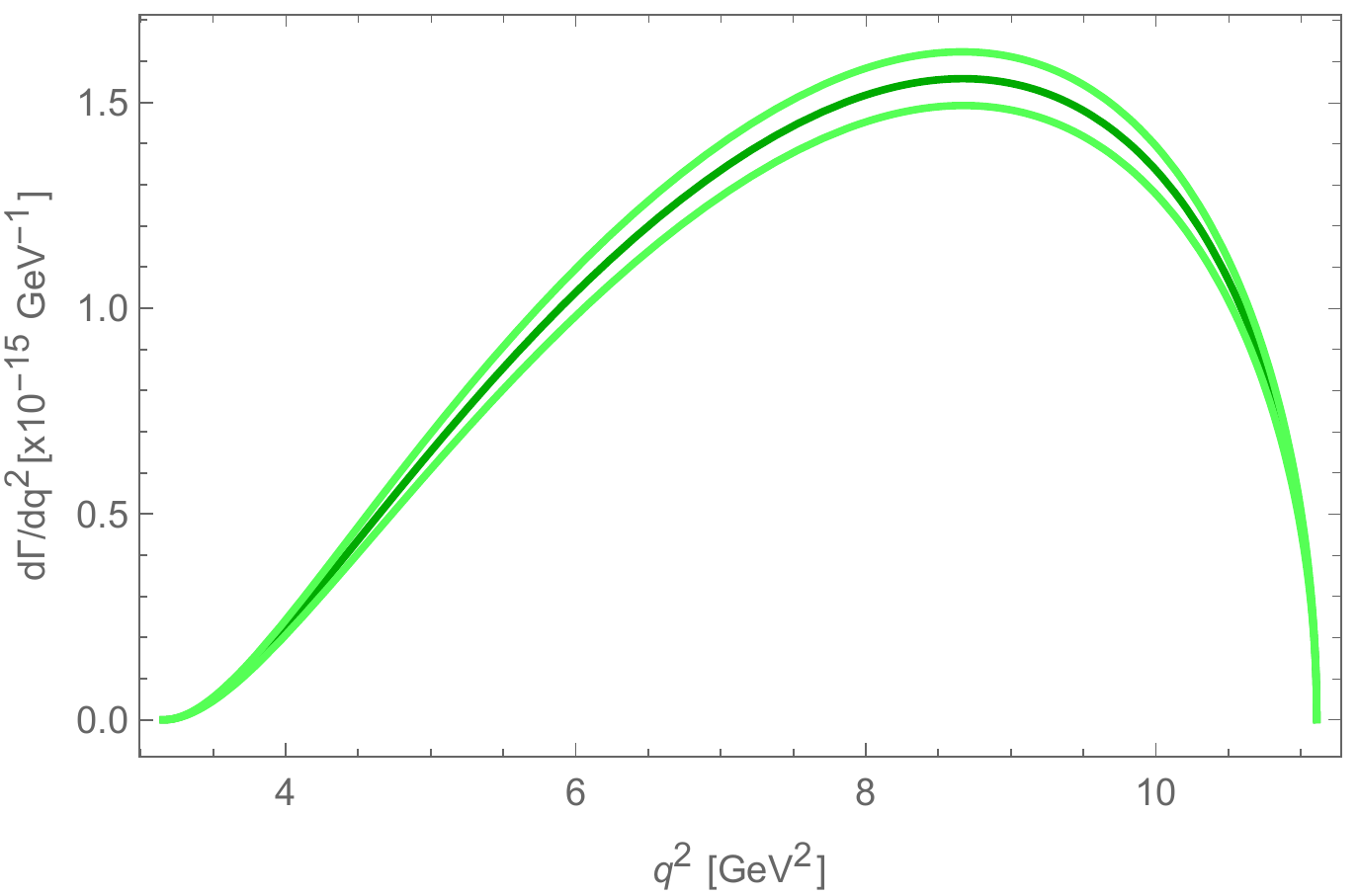}~~~
 \includegraphics[scale=0.55]{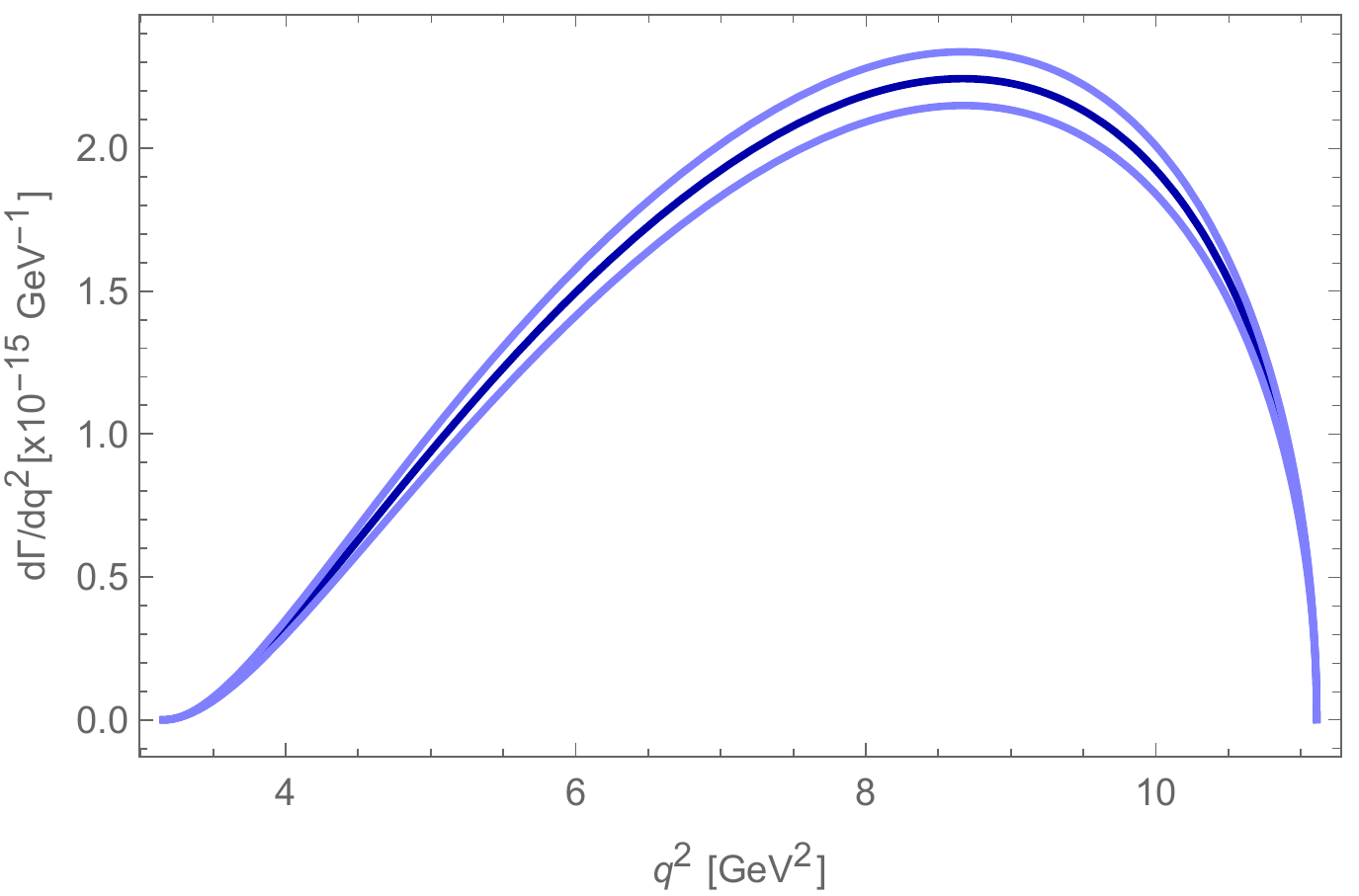}~~~
 \caption{The differential cross section distribution from the Mathematica code. The left panel is the Standard  Model results and the right panel is the new physics results at $g_s= g_p= g_R= g_T=0$ and $g_L=-2.2$.}
 \label{figure2}
\end{figure}

\begin{figure}[!h]
% \centering
\hspace*{-1cm} 
 \includegraphics[scale=0.33]{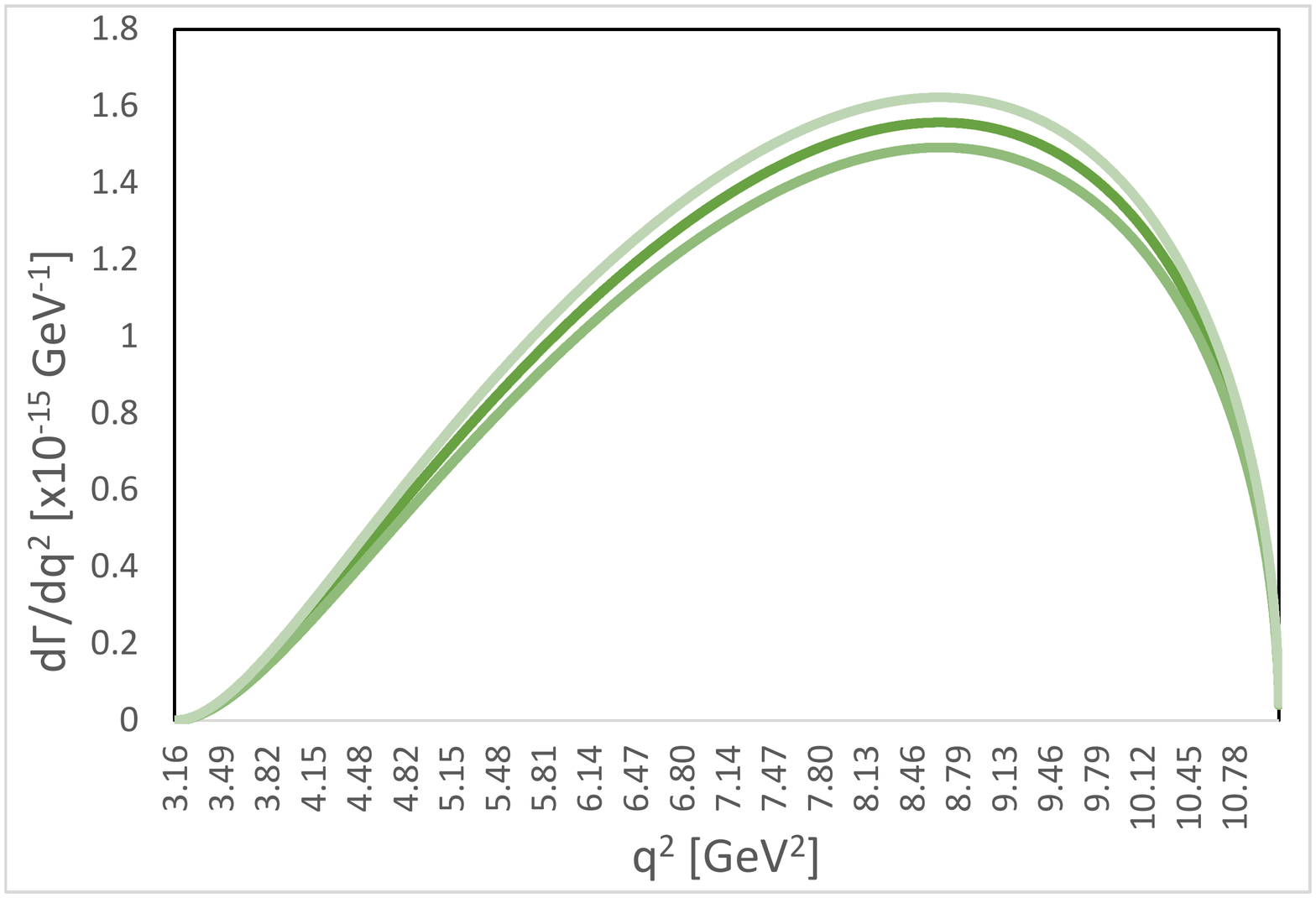}~
 \includegraphics[scale=0.33]{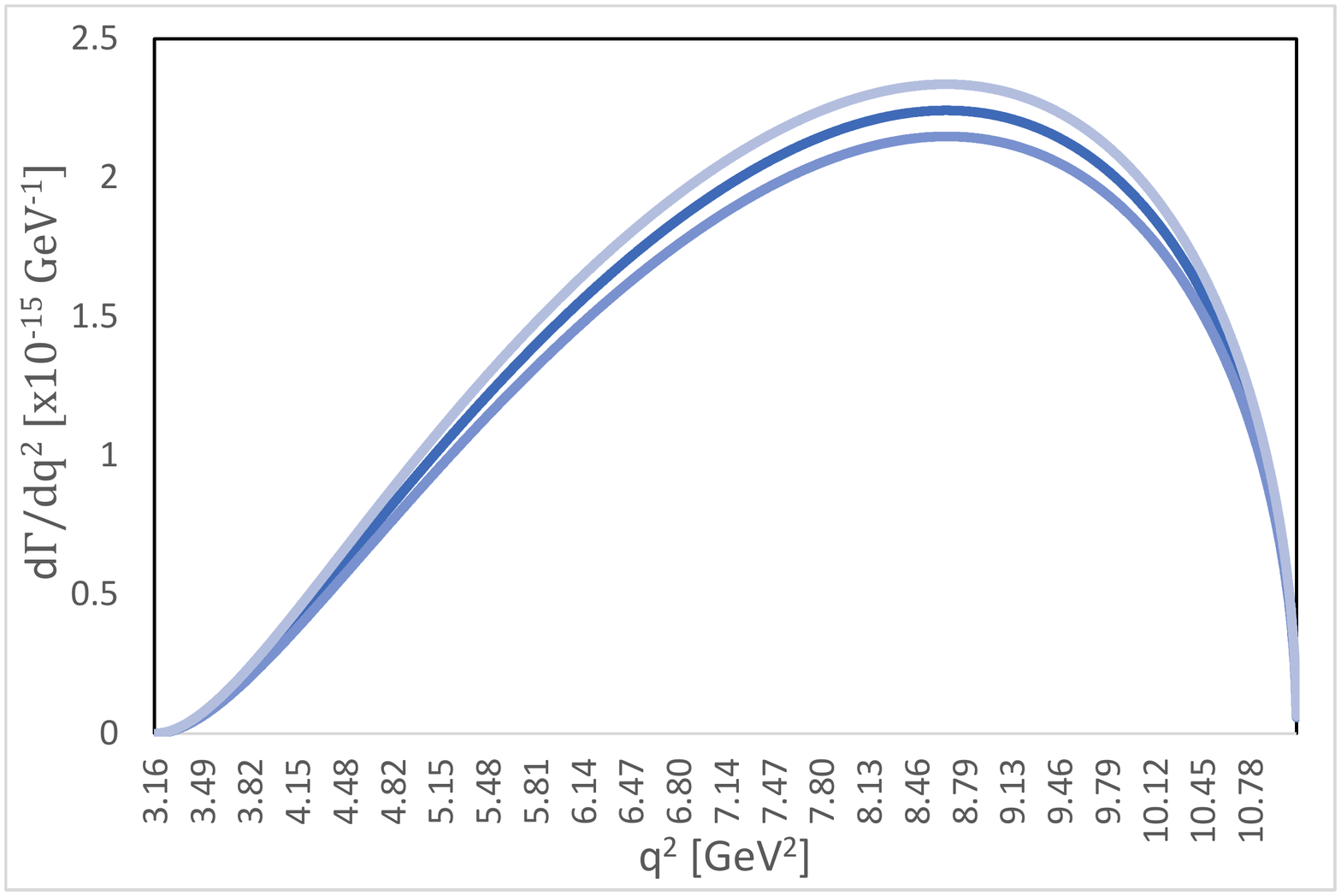}~
 \caption{The differential cross section distribution from the c-code. The left panel is the Standard  Model results and the right panel is the new physics results at $g_s= g_p= g_R= g_T=0$ and $g_L=-2.2$.}
 \label{figure3}
\end{figure}

Once the results are stored in the file, the data can be used in a graphing software to create the graph of the decay rate and its error. The format of the data file is shown in Fig.~\ref{figure1}. The resulting graphs of the differential distribution from both the Mathemtica and c-codes are shown in Figs.~(\ref{figure2}, \ref{figure3}). The graphs show that the c-code results match the Mathematica results and agree with the results in Ref.~\cite{Datta:2017aue}. This is a verification that the c-code runs correctly. The results of the speed test for the c-code and Mathematica code are presented in Fig.~4 and Table~1. We have run both codes for 10, 100, 1000, 10000 data points for five trials in each case. We have recorded the wall clock time for each trial and averaged them. The results show that the graph of the wall clock time versus the number of data points for both the c-code and Mathematica code are linear. By computing the ratio of the slopes, we have concluded that the ratio of the wall clock time for the c-code compared to the Mathematica code is 1:64.2 per data point.

\pagebreak

%\begin{figure}[!h]
% \centering
% \includegraphics[scale=0.9]{C-SM.png}~
% \includegraphics[scale=0.9]{C-NP.png}~
% \caption{The differential cross section distribution from the c-code. The left panel is the Standard  Model results and the right panel is the new physics results at $g_s= g_p= g_R= g_T=0$ and $g_L=-2.2$.}
% \label{figure3}
%\end{figure}

\begin{figure}[!h]
 \centering
 \includegraphics[scale=0.43]{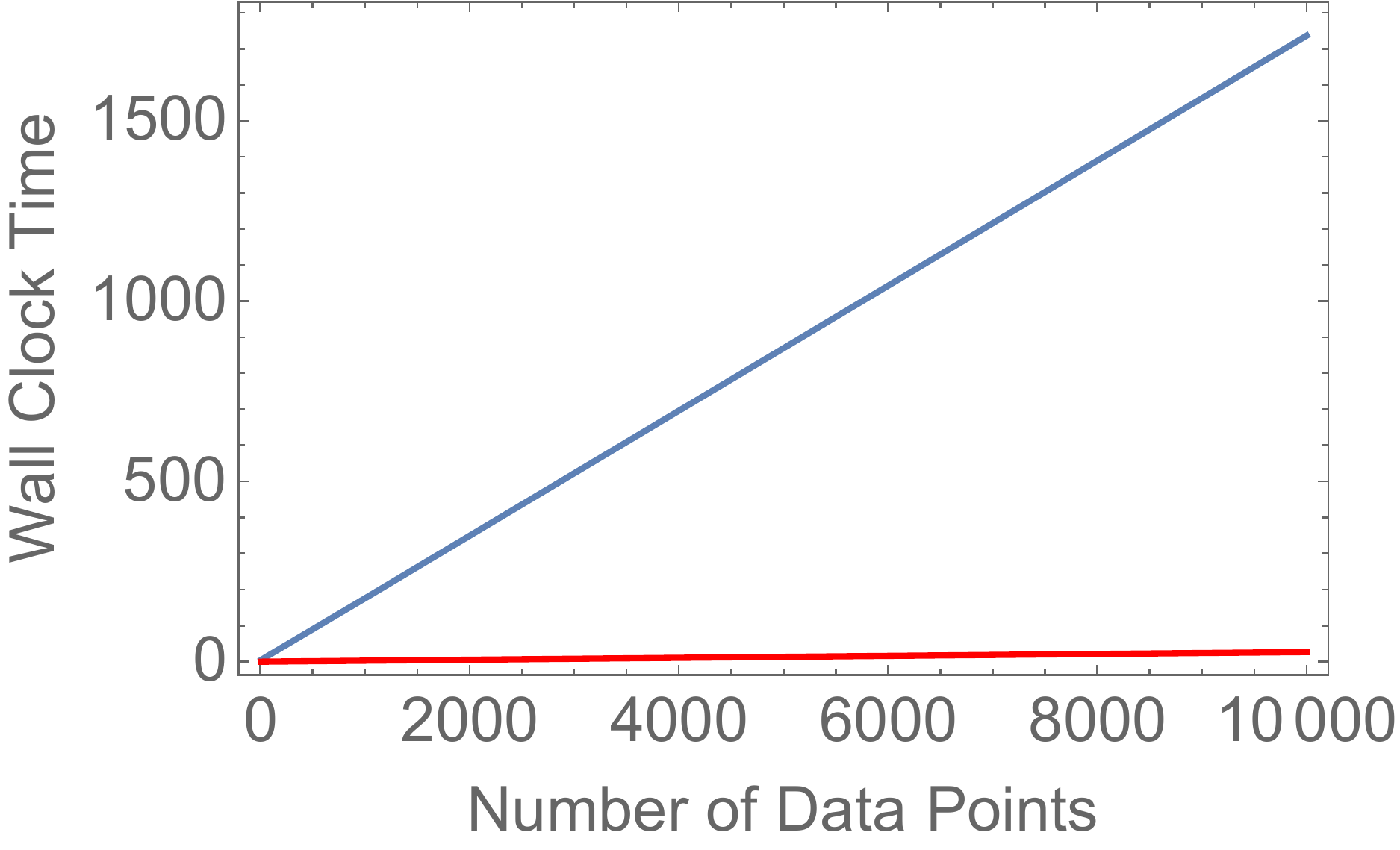}~~~
  \includegraphics[scale=0.43]{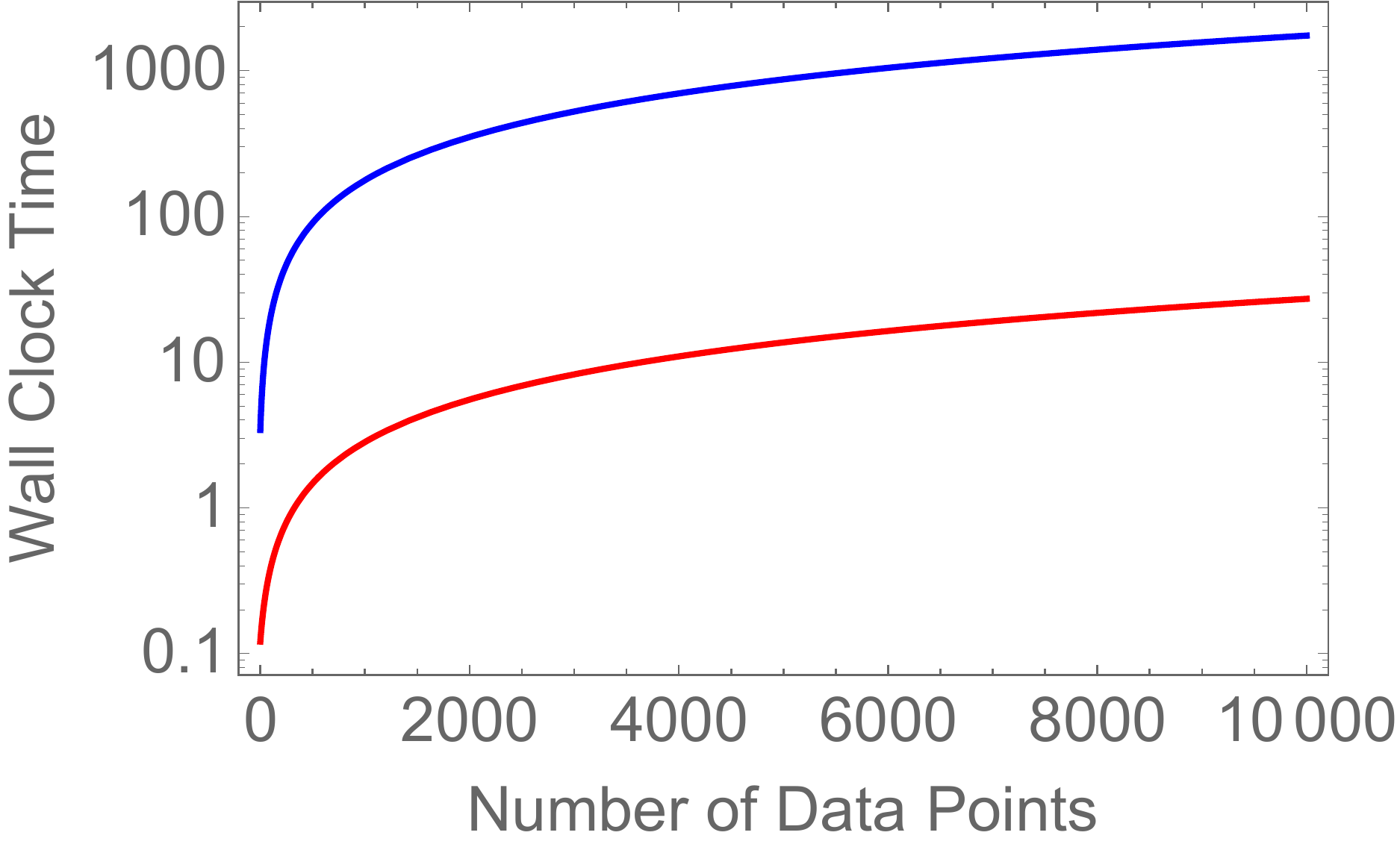}~~~
 \caption{Results of of the speed test. This shows that the c-code (red) performs faster than the Mathematica code (blue). Left and right panels are the linear and log plots, consequently, of the the wall clock time versus the number of data points.}
 \label{figure4}
\end{figure}

\begin{table}[!h]
 \begin{tabular}{||c | c | c ||} 
 \hline
 Number of points & Average Time of Mathematica Code & Average Time of C Code   \\ [0.5ex] 
 \hline\hline
 10 & 5.4264  &0.1076   \\ 
 \hline
 100  & 21.2736 & 0.3652  \\
 \hline
 1000 & 175.7828 & 2.8656  \\
 \hline
  10000 & 1736.1958 & 26.8708   \\ 
  \hline
\end{tabular}
\caption{Results of of the speed test for the c-code and the Mathematica code. }
\label{Table1}
\end{table}

\pagebreak

\section{Future of Code}
In the future of the code, a Graphical User Interface (GUI) will be made for easier use. The GUI will include the ability to select the desired file from the computer's file explorer, the ability to select the values of all variables, and the ability to change the number of data points created.  These functions will expand the capability of the code without having to go in and edit it directly, as well as create ease of use.
Another plan for the code is to expand its scope to create data files for the higher order decay rates, for it only gives data for the nominal, as well as other desirable elements the code calculates.  It will also be expanded to solve similar problems.

\section{Conclusion}

The hadronic transition of the $\Lambda_b \rightarrow \Lambda_c$ can be parametrized in terms of scalar, vector, and tensor QCD lattice form factors. In Ref.~\cite{Datta:2017aue}, one of the authors (SM) has written a Mathematica code to read the form factor data files. In this paper, we wrote a c-code to read the QCD form factors data files and generate data points for the decay rate of the process $\Lambda_b \to \Lambda_c \,\tau^- \bar{\nu}_\tau$. The c-code has provided the same results as the Mathematica code as a verification of the c-code. The speed test for the Mathemtica code and c-code showed that the c-code found to be faster and smaller than the Mathematica code. The ratio of the wall clock time spent by the c-code to the Mathematica code is 1:64.2 per data point.  The space taken up by the c-code is 26 KB while the Mathematica code takes up 107 KB.

\bigskip
\noindent
{\bf Acknowledgments}:
We thank Stefan Meinel for providing the Mathematica code that reads the data file of the QCD lattice form factor. B.S. acknowledges the hospitality of Stefan Meinel at the Department of Physics, University of Arizona, where he gave B.S. a lesson on the mathematics and physics behind the code, as well as the algorithm of the Mathematica code.  This work was financially supported by the Student/Faculty Research Engagement (SFRE) Grant and Summer Undergrad Research Experience (SURE) Grant (B.S.).

\pagebreak

%%%%%%%%%%%%%%%%%%%%%%%%%%%%%%%%

\end{document}